\documentstyle[prl,aps,psfig]{revtex}

\begin{document} 
  
\draft  

\twocolumn[ 
\begin{center}
{\bf Valence band photoemission from the GaN(0001) surface}\\ 
{T. Strasser, C. Solterbeck, F. Starrost, and W. Schattke}\\  
{\it Institut f\"ur Theoretische Physik und Astrophysik, Christian-Albrechts-Universit\"at, 
Leibnizstr. 15,}\\
{\it D-24098 Kiel, Germany}\\
{submitted to Phys. Rev. B, 17.05.1999}\\
\end{center}]
 
 
%
 
\begin{abstract} 

A detailed investigation by one-step photoemission calculations
of the GaN(0001)-(1x1) surface 
in comparison with recent experiments is presented in order to clarify 
its structural properties and electronic structure. 
The discussion of normal and off-normal spectra reveals through the
identified surface states clear fingerprints for the applicability of a
surface model proposed by Smith {\it et al.} \cite{SmNo97}. Especially 
the predicted metallic bonds are confirmed. 
In the context of direct transitions the calculated
spectra allow to determine the valence band width and to argue in
favor of one of two theoretical bulk band structures. Furthermore a commonly 
used experimental method to fix the valence band maximum
is critically tested. 

\end{abstract}  
 
%
\pacs{79.60.-i, 73.20.At, 73.61.Ey}  

\section{Introduction} 
 
The wide band gap semiconductor GaN has experienced exciting applications in blue light  
emitting diodes and laser diodes. For a further improvement of the quality of the material a  
better understanding of the structural and electronic properties is necessary.
The energetic positions of critical points differ by 0.8 eV for the
existing band structure calculations of wurtzite GaN \cite{RuCo93,VoPo96,YeLi98}. 
Furthermore, the  geometrical structure of  
the GaN(0001) surface is still being debated \cite{RaBe97,SmNo97,FrPa98}. 
 
The most powerful tool for examining the electronic structure of
semiconductors is the angle  
resolved ultraviolet photoemission spectroscopy (ARUPS). The spectra give insight into the  
valence band structure of the bulk as well as the surface. Besides the part
of direct interest, i.e. the initial bound state, the  
photoemission process also involves the excitation to outgoing scattering states with the transition 
probability given by matrix elements. Therefore, as already demonstrated for the cubic  
GaN(001) surface \cite{StSc97}, a full account of the experimental data can only be attained  
by a comparison with photocurrents calculated within the one-step model. 
 
As a starting point we use a GaN(0001)-(1x1):Ga surface, as it is predicted by total energy  
calculations performed within the local density formalism \cite{SmNo97}. The calculated  
photoemission spectra in normal emission are examined with respect to contributions from the  
bulk band structure as well as from surface states. For example, we identify a structure near the  
lower valence band edge as resulting from a surface state. Only by taking this state into  
account, the correct energetic position of the band edge can be extracted. Based on
the  
detailed understanding of the photocurrent the flexibility of the calculation
allows us to   
adjust the underlying bulk band structure to the experimental results. This means
that we are able to correct the position of the valence band maximum, which is an  
important value for determining band offsets and band bending. Though abandoning
parameter-free                                                      
modeling thereby, one gains experience how peaks are shifted and intensities are deformed by the  
matrix elements. The true position of the bands can be much better
determined in such an interpretation of  
experiment than using standard band mapping methods. 
 
Off-normal photoemission spectra provide an enhanced surface sensitivity.  
Near the upper valence band edge, experiment has pointed out an $sp_z$ orbital related surface  
state \cite{DhSm97}. Identifying this state, which depends sensitively on
the surface geometry, in the theoretical spectra, we can  
connect the geometric and electronic structure with the measured photocurrents. 
 
This paper is organized as follows. First a short overview about the theory is
given, followed by   
a detailed analysis of the initial surface band structure and the used final bands. 
Then the results for normal-emission spectroscopy 
from the GaN(0001)-(1x1):Ga surface are presented, together with a detailed interpretation in  
comparison with experiment. It is shown how the theoretical band structure calculation can  
be related to experiment and how uncertainties in the experimental interpretation can be  
removed. Finally, we present the results for off-normal emission, comparing with 
experimental data, too. 
 
\section{Theory}  
In this section, we briefly discuss the theoretical techniques used in our calculation of the  
photocurrent. For details see the references 
\cite{HeSc89,HeSc93}. 
 
We calculate the photocurrent within the one-step model. The photocurrent $I$ is given by: 
 
\begin{eqnarray}  
I\; &\sim\ &\;\sum_{i,j}\; \langle 
\Phi_{LEED}^\ast(E_{fin},k_\|)| {\bf 
A_{0}}\cdot {\bf p} | \Psi_i \rangle \nonumber \\ 
&&\; G_{i,j}(E_{fin}-h\nu,k_\|) \langle \Psi_j | {\bf p}\cdot{\bf A_0}| 
\Phi_{LEED}^\ast(E_{fin},k_\|) \rangle  
\end{eqnarray} 
 
\begin{figure}[h]\centering 
\centerline{\psfig{figure=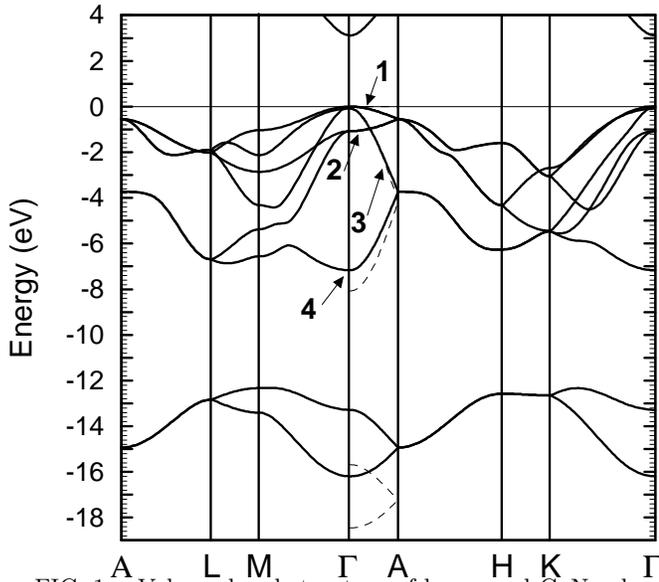,width=8.6cm}} 
 \caption{\label{ganvolumen} Valence band structure of hexagonal GaN 
  calculated with EHT-parametrization according to 
  Table I (solid line: set A, dashed line: set B).}    
\end{figure}

For simplicity the vector potential ${\bf A_0}$ is kept constant. $G_{i,j}$ represents the 
halfspace Green's  
function of the valence states, given in a layer-resolved LCAO basis $\Psi_i$ . Our basis set  
consists of the 4$s$ and 4$p$ atomic orbitals of gallium and the 2$s$ and
2$p$ atomic orbitals of  
nitrogen, taking into account the coupling up to fourth nearest neighbor
atoms.  
The Hamilton matrix is calculated according to the Extended-H\"uckel-Theory  
(EHT). Its parameters, employed for bulk and halfspace calculations,
are adjusted to published {\it ab-initio} bulk band structures using a 
genetic algorithm \cite{StSc96}.
We use two different sets for the parameters. One set is  
adjusted to the GW quasiparticle band structure of Rubio {\it et al.} \cite{RuCo93}. The other set  
is adjusted to a self-interaction and relaxation corrected  pseudopotential band structure  
calculation by Vogel {\it et al.} \cite{VoPo96}. The two sets of parameters belonging to
these band  
structures are presented in Table \ref{eht}. Figure \ref{ganvolumen} shows the resulting bulk band  
structure according to Vogel {\it et al.} (solid lines). Along $\overline{\Gamma A}$, also the band  
structure adjusted to Rubio {\it et al.} is shown (dashed lines). The main difference is the energetic  
position of the lower valence band edge at $\overline{\Gamma}$ near -8.0 eV,
where the calculations differ by nearly 0.8 eV. 
 
The electronic structure of the surface is determined by the calculation of the ${\bf k_{\|}}$- 
resolved density of states (DOS) from the halfspace Green's matrix
$G_{i,j}$, the same as used for the photocurrent. It takes into account relaxation and  
reconstruction at the surface. The resolution of the DOS with respect to atomic layers and  
orbital composition allows for a detailed characterization of the bands and their corresponding 
photocurrents. 
 
The final state of photoemission is a scattering state with asymptotic boundary  
conditions. For a clean surface its wave function is determined by matching  
the solution of the complex bulk band structure to the vacuum solution, representing the  
surface by a step potential \cite{Br89,Pe69}. This treatment is best suited for discussing the  
photoemission peaks in terms of direct transitions with conservation of the surface  
perpendicular wave vector $k_{\perp}$ since the final state is described
inside the crystal as a sum over bulk solution of different $k_{\perp}$.  
These solutions of the complex bulk bands are   
calculated with an empirical pseudopotential. For GaN we use the pseudopotential 
form factors of Bloom {\it et al.}  
\cite{BlOr74}. The damping of the wave function inside the crystal is described by the 
imaginary part of an optical potential. 
 
In Eq. (1), the transition matrix elements $\langle \Phi_{LEED}^\ast(E_{fin},k_\|)| {\bf 
A_0} \cdot {\bf p} | \Psi_i \rangle$ between the final state and layer Bloch sums are  
numerically integrated in real space. 
Their dependence on atomic layers and orbitals permits a detailed analysis
of the spectra.
 
\section{results and discussion} 
\subsection{Electronic structure} 
 
\begin{figure}[h]\centering  
\centerline{\psfig{figure=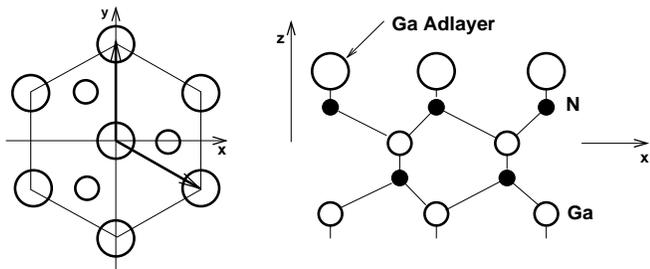,width=8.6cm}}        
 \caption{\label{ganhexgeo} The geometry of the GaN(0001)-(1x1):Ga 
  surface according to \protect\cite{SmNo97}.}    
\end{figure}

In this section we discuss our results for the electronic structure of the  
GaN(0001) surface. We use a surface geometry of Smith
{\it et al.} \cite{SmNo97} (shown in Fig. \ref{ganhexgeo}),
derived from total energy calculation and STM examinations.  Directly atop the  
nitrogen atoms sits a full monolayer of gallium adatoms, forming a (1x1)
surface. In Fig. \ref{ganoberband} the surface band  
structure is presented, calculated with the parameters of set A in Table I. 
The bands are determined from the peaks in the ${\bf k_{\|}}$ resolved DOS
of the four topmost atomic layers.  

\begin{figure}\centering  
\centerline{\psfig{figure=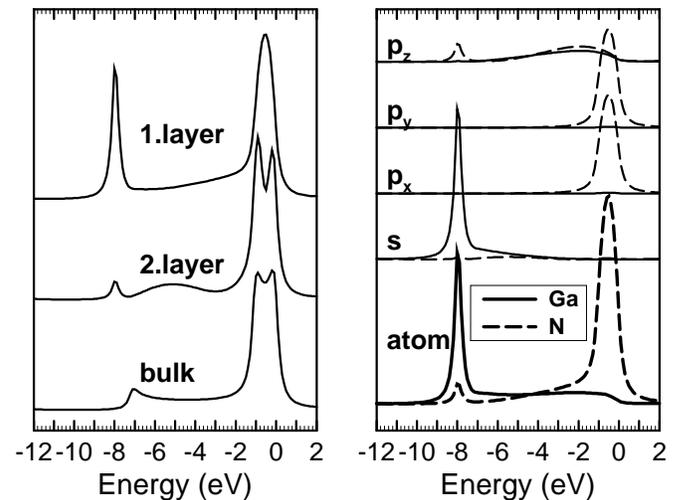,width=8.6cm}}        
 \caption{\label{gandos} Density of states at $\overline{\Gamma}$ for  
 GaN(0001)-(1x1):Ga. On the left hand side the layer resolved DOS is shown, 
 while on the right the DOS of the first layer is further resolved into its 
 atomic and orbital contributions. Calculated with the parameters from set A (see 
 Table I).}     
\end{figure}

Within the fundamental gap there are two surface states, labeled (a) and (b). They  
can be identified as $p_x$ and $p_y$ derived bridge bonds between the Ga adlayer atoms.
Smith {\it et al.} found these strongly dispersive metallic bonds to be 
responsible for the stability of this surface \cite{SmNo97}. 
The state (c) is built up by the Ga $s$ and $p_z$ orbitals, in
$\overline{K}$ and $\overline{M}$ with strong contributions from the underlying  
N $p_z$ orbitals. Along $\overline{\Gamma M}$ and mostly along 
$\overline{\Gamma K}$ the state (c) is in resonance with the bulk, mixing with the  
nitrogen $p_x$ and $p_y$ orbitals. Especially in $\overline{\Gamma}$, these orbitals exhibit a  
strong contribution to the density of states, as can be seen in Fig. \ref{gandos}.  

Along $\overline{\Gamma M}$ the surface  
resonances (e) and (f) are made up by the nitrogen $p_x$ orbitals with small contributions from  
the underlying Ga $s$ and $p_x$ orbitals. For band (d) we find strong contributions from the  
nitrogen $p_y$ orbitals and from the $p_y$ orbitals of Ga lying below the
nitrogen layer. Along $\overline{\Gamma  
K}$ we find similar resonances (h, i, k), which can be resolved into contributions from the N  
$p_y$ orbitals with a smaller amount from the N $p_x$ and the subsurface Ga $p_x$ and $p_y$  
orbitals. The band (g), seen at -4.0 eV near $\overline{\Gamma}$ is a nearly invisible 
structure in the density of states (Fig. \ref{gandos}), built up by broad contributions  
from nitrogen and gallium $p_z$ orbitals. In the heteropolar gap we find a strong Ga $s$  
surface state (l) (see also Fig. \ref{gandos}) located at -8.0 eV clearly below the  
lower bulk band edge. As can be seen in the DOS, this state contains also contributions from  
the N $p_z$ orbitals. Along $\overline{\Gamma M}$ and  $\overline{\Gamma K}$ this band 
shows a strong dispersion towards lower binding energy, becoming a surface  resonance 
between $\overline{M}$ and $\overline{K}$. Near the lower valence band edge we find a  
second resonance (m)  formed by the N $p_x$ orbitals and Ga $s$ orbitals  
from deeper atomic layers. In $\overline{K}$ we find also contributions from the N $p_y$  
orbitals. 
Altogether, taking into account the states inside  
the fundamental gap and the surface state in the heteropolar gap, the surface band structure  
of the GaN(0001)-(1x1):Ga surface shows a similar behavior as the cubic GaN(001)-(1x1):Ga  
surface \cite{NeWa98,StSc97}. 
 
\begin{figure}\centering  
\centerline{\psfig{figure=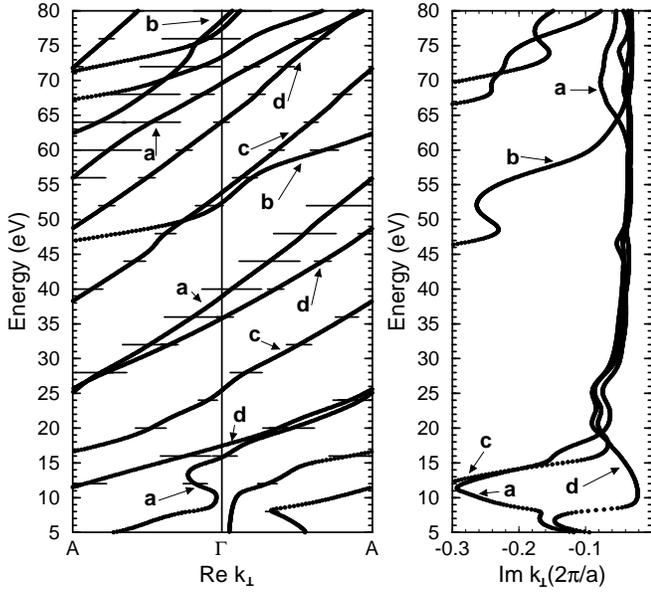,width=8.6cm}} 
 \caption{\label{gankomplexe} Complex band structure of GaN for the 
 symmetry line $\Delta$. Bars indicate the magnitude of the expansion 
 coefficients of the final state with respect to the complex bulk bands.}     
\end{figure}

The complex final bands are shown in Fig. \ref{gankomplexe} . They are calculated along the  
symmetry line $\Delta$, which corresponds to normal emission. Since we introduced
an imaginary optical potential all bands are damped. The horizontal bars denote the weight which  
single complex bands carry in the final state. With respect to this
criterion, only the four most important
bands for the  
photoemission are shown. They have strong contributions to the final state  
(labeled (a), (b), (c), and (d)). Below 15 eV final state energy state (d) is the most  
important one. In this energy range state (a) and (c) reveal large imaginary parts, being  
responsible for a strong damping of these states inside the crystal. Between 15 and 65 eV 
state (a) contributes dominantly. Above 65 eV band (b) yields the essential contribution to the  
final state.

\subsection{Normal Emission} 
 
\begin{figure}\centering 
\centerline{\psfig{figure=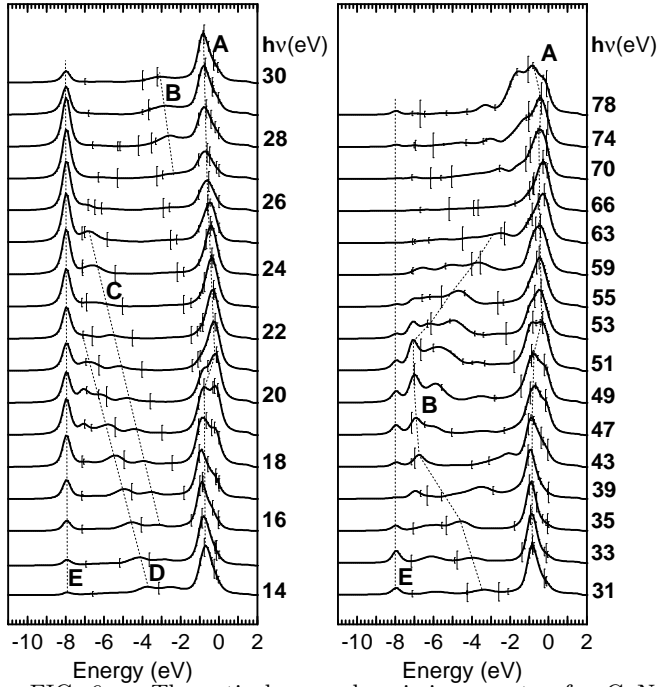,width=8.6cm}}    
 \caption{\label{gannorppolxz} Theoretical normal emission spectra for  
 GaN (0001)-(1x1):Ga. We have normalized each spectrum separately
 to an equal amount in the maximum of the 
 photocurrent. The bars indicate binding energies at which direct 
 transitions would be positioned. Their dispersion is drawn by dotted lines as a guide the eye.  
 The energy zero is the VBM, the light is chosen incident along the $xz$ plane 
 with $p$ polarisation. The spectra are calculated  from set A in Table (I).} 
\end{figure}

\begin{figure}\centering 
\centerline{\psfig{figure=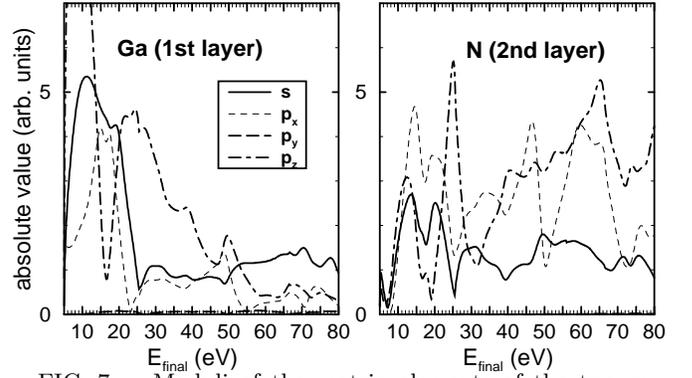,width=8.6cm}} 
 \caption{\label{gannormat} Moduli of the matrix elements of the two  
 uppermost atomic layers of the GaN(0001)-(1x1):Ga surface resolved into 
 orbitals. The matrix elements are calculated for $p$-polarized light incident 
 in the $xz$ plane.} 
\end{figure}

Figure \ref{gannorppolxz} presents normal emission spectra for the GaN(0001)-(1x1):Ga surface,  
which are  
calculated for photon energies from 14 up to 78 eV. The radiation is chosen to be incident  
within the $xz$ plane at an angle of $45^\circ$ to the surface normal. The radiation is
polarized parallel to the plane of incidence. The spectra are dominated by two structures, near -1.0 (A)  
and -8.0 eV (E) . The last one coincides with the strong Ga $s$ surface state near -8.0 eV,  
which can be seen in the DOS at $\overline{\Gamma}$ in Fig. \ref{gandos} . The emission  
from this peak is localized in the topmost gallium layer, which can be
proven by analyzing the  
orbital and layer resolved matrix elements, as presented in Fig. \ref{gannormat}. For final state  
energies between 8 and 21 eV the $s$ orbitals of the gallium adlayer atoms show
large matrix elements. This agrees with the strong emissions at -8 eV for photon  
energies between 16 and 29 eV. For higher final state energies the Ga $s$ 
matrix elements as well as the peak heights are much smaller. At 25 eV and above 40 eV final state  
energy the matrix elements of 
the first nitrogen $p_z$ orbitals become appreciable. So the emissions near 33, 49 eV and 78 eV  
photon energy are enhanced by emissions from nitrogen $p_z$-orbitals,
although they show much weaker DOS at $\overline{\Gamma}$ near -8.0 eV than the gallium $s$ orbitals.  
 
The leading peak near -1.0 eV valence energy (A) is connected to the high and broad    
density of states (Fig. \ref{gandos}), resulting from the nitrogen $p_x$- and $p_y$ orbitals. 
While emissions from the $p_y$ orbitals are forbidden by selection rules, the nitrogen $p_x$  
matrix elements exhibit minima near 25, 50 and 73 eV final state energy. This is  
consistent with the decreasing intensity of the leading peak near 27, 51 and  74 eV  
photon energy whereby the behavior at 27 eV is especially convincing. 
For 51 eV the smaller  
intensity of the leading peak is one reason for the more  
pronounced intensity of those at higher binding energies, because in Fig.
\ref{gannorppolxz} each spectrum is normalized seperately to an equal amount in
the highest peak.  
Furthermore, for analyzing structure  
(A) we have to take into account direct transitions assuming exact  
conservation of the perpendicular wave vector. For a given  
excitation energy we determine the binding energies at which transitions
from the initial bulk band structure into 
the complex final band structure are possible. These binding energies are
plotted in the 
photoemission spectra with bars whose length indicates the contributions of the complex band to  
the final state. For structure A we have to consider the initial bands 1, 2 and 3 (see  
Fig. \ref{ganvolumen}). Hints for the contribution of direct transitions to
structure (A) are 
the dispersion for photon energies between 14 and 20 eV at the lower 
binding energy side (from initial state (2) into final band (d)) and between 
20 and 47 eV photon energy were the leading peak disperses from -0.1 eV 
to -1.1 eV (initial band (1) and (2) into final band (a)). 
For photon energies of 17 eV (final band (d)), 39 eV (final band (a)) and 74 
eV (final band (b)) shoulders from direct transitions from the VBM can be 
seen. 
 
Besides the two prominent structures there exist weaker intensities with
strong dispersion. (C) and (D) can  
be identified as emissions from the initial bands (3) and (4) into the final bands (d) and
(c), respectively.  
Between 28 and 63 eV photon energy we find the dispersive structure (B). It can be explained by  
direct transitions from the valence band (4) into the final band (a).
At 47 eV photon energy the structure reaches the  
highest binding energy with -7.2 eV. Near the lower valence band edge above (E) and interfering  
with (B) there are further weak structures in the photon range from 39 to 59 eV which result from  
nitrogen $p_z$ orbitals. Near -5.5 eV emissions from the nitrogen $p_z$ orbitals  
from the second  
nitrogen layer arise for photon energies between 49 and 59 eV. Emissions from the third  
nitrogen layer $p_z$, located at -6.7 and -3.2 eV are visible in the photon range (51,55) eV  
and (39,49) eV respectively.  
 
Compared with the GaN(001) surface \cite{StSc97}, direct transitions are less significant for  
wurtzite GaN in the range up to 78 eV photon energy. Further calculation shows, that above 98  
eV photon energy a strong dispersive structure belonging to the initial state band (4) and the  
final state (b) appears, reaching its highest binding energy near 118 eV photon energy.    
 
\subsection{Normal emission - comparison with experiment} 
 
In this section we compare our calculated spectra with experimental results in normal emission  
performed by Dhesi {\it et al.} \cite{DhSm97} for photon energies between 31 and 78 eV. The  
measurements were done at an wurtzite GaN film, grown by electron cyclotron resonance  
assisted molecular beam epitaxy on sapphire substrates with subsequent annealing. The  
spectra were detected with synchrotron radiation, incident at $45^\circ$ to the surface normal.

\begin{figure}\centering 
\centerline{\psfig{figure=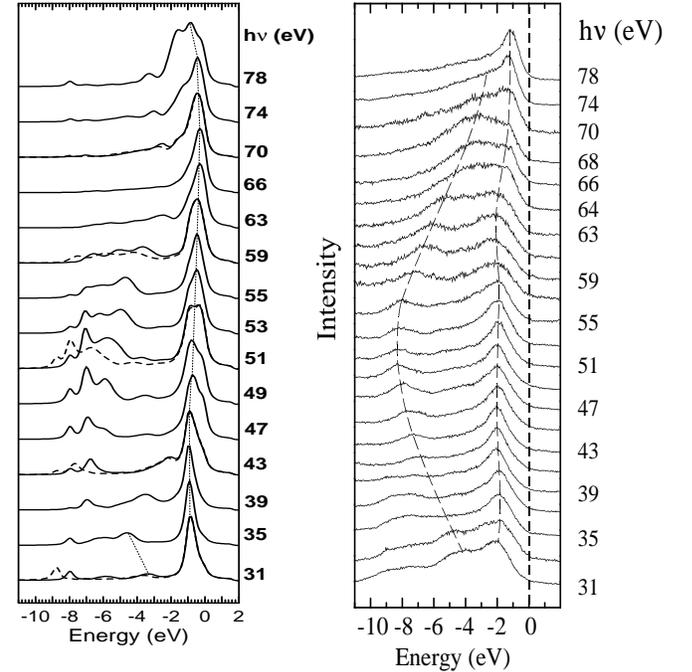,width=8.6cm}} 
 \caption{\label{gannortheoexp} Comparison between  theoretical (left hand side) 
 and experimental photoemission spectra of Dhesi {\it et al.} 
 \protect\cite{DhSm97} in normal emission. The theoretical spectra are 
 calculated from set A (see Table I), except for the dashed spectra, which are 
 calculated from set B.} 
\end{figure}

Figure \ref{gannortheoexp} shows on the right hand side the experimental
results. Energy zero is the VBM, which was determined from the spectra
by extrapolating the leading edge. In comparing the spectra we will show that this
technique places the experimental VBM 1.0 eV above the VBM as taken 
from the band structure. 
  
On the left hand side of Fig. \ref{gannortheoexp} our theoretical results, as discussed in  
the section (B), are plotted (solid lines). The experimental data show a dominant structure near
-2.0 eV which can be associated with the theoretical peak at -1.0 eV. Like the  
theoretical results, this structure exhibits some dispersion to lower binding energies between 63  
and 66 eV photon energy (theory between 59 and 63 eV). These emissions can be explained  
by transitions from the N $p_x$ orbitals and by direct transitions from the bulk bands  
(1) and (2).
Furthermore, the experimental results reveal a dispersing structure between -4.2 and -8.2
eV, which is also visible in the theoretical results (between -3.2 and -7.2 eV). In both series, the
emissions from that structure become weak for photon energies around 39 eV. Near 55 eV
photon energy both experiment and theory show enhanced emissions, dispersing back to lower
binding energies. Around 66 eV photon energy the emissions near -4.0 eV are much weaker in
theory than in experiment. The theoretical spectra show a significant doubling of the leading
peak at h$\nu$=78 eV, with emissions near -0.8 eV and -1.8 eV. A similar
effect is not seen in the experimental results of Dhesi {\it et al}.
However, recent measurements by Ding {\it et al.} display  
the double maximum \cite{DiHo}. The latter experiment was performed on a 
GaN(0001)-(1x1) surface, also grown on sapphire but by means of metal organic chemical vapour 
deposition (MOCVD). For photon energies of 75 and 80 eV the experimental 
spectra by Ding {\it et al.} show peaks near the upper valence band edge and near -2.0 eV, 
which can be connected to the peaks in the theoretical spectra for 74 and 78 eV photon energy. 

All over all, we can identify two significant structures from the experimental
data by Dhesi {\it et al.} in our calculated spectra. Comparing their
energetical position we recognize, that the theoretical structures are at 1.0 eV
lower binding energy. We explain this difference by an inaccuracy of the
experimentally determined VBM of 1.0 eV. It should be pointed out, that this
error also explains the energetical shift of 1.0 eV which is necessary to
match the experimental band structure of Dhesi {\it et al.}  with the theoretical 
band structure in Ref. \cite{DhSm97}.

Apart from the two discussed series, Fig. \ref{gannortheoexp} includes some dashed
lined theoretical spectra. These spectra are calculated with the EHT
parameters of set B, see Table \ref{eht} . The parameters are  related to the band structure of Rubio 
{\it et al.} with a  
valence band width of 8.0 eV. The spectra are similar to the calculated results already discussed. 
The leading
peak is almost unchanged, while the emissions near the lower valence band edge are shifted by    
0.8 eV. This statement is true for the whole theoretical series calculated with the  
parameters of set B. Comparing with experiment, we can point out two results. The leading
peak between -1.0 and -0.3 eV in both theoretical series can be  
identified with the experimental structure between -2.0 and -1.2 eV. This
underlines that the experimental VBM has to be shifted by nearly 1.0 eV to higher binding energies
as already stated.  
Similarly the emissions near -8.2 eV in the experimental spectra can be assumed to lie at -7.2
eV, which would be consistent with the theoretical spectra calculated by
the band structure of Vogel {\it et al.} \cite{VoPo96} (parameters of set A).
This means that we are able to determine the valence band width to 7.2 eV,
by comparing the experimental spectra with calculated photocurrents based
on different band structures.

Furthermore, in the experimental paper of Dhesi {\it et al.} it is pointed out that at lower photon  
energies a non-dispersive feature with a binding energy of approximately  -8.0 eV is visible  
\cite{DhSm97}. In Ref. \cite{DhSm97} the peak is explained with final state or density of state effects, 
rather than with a surface state. Our examination however, for photon energies lower than 30 eV clearly reveals  
significant emissions from a gallium $s$ surface state at that energy, and not from the band  
edge (see section (B)). The theoretical band edge is found at 0.8 eV lower binding energy, and additionally,
the variation of the intensity with photon energy is associated with the matrix elements and uniquely attributes  
this emission to a surface state.
In both the theoretical and experimental spectra weak emissions for photon
energies between 31
and 39 eV around the lower valence band edge are observed which can be additionally
attributed to the surface band emission (see Fig. \ref{gannortheoexp}).   
Near -8.0 eV the experiment shows enhanced emissions for photon energies
between 47 and 55 eV , which are also seen in the theory around -7.2 eV.
The experimental peaks at highest binding energy are broad enough for
including also the emissions from the surface state near -8.0 eV in theory
(gallium $s$ and nitrogen $p_z$ related).
The association of a non-dispersive structure with a theoretically estimated band edge merely because of its
energetical vicinity can easily lead to erroneous band mapping \cite{StSc97},
especially as surface states often develop near band edges. 

Summarizing the above discussion, we have demonstrated that 
there is a clear aggrement between the theoretical and
experimental results for a wide energy range. 
This aggrement allows us to show that the determination of
the VBM by extrapolating the leading edge could involve significant errors. 
The determination of the VBM is an important step in the investigation of band bending 
and valence band discontinuity in  
heterojunctions \cite{DiHo97,WuKa98}. Especially, Wu {\it et al.}\cite{WuKa98} investigated the  
band bending and the work function of wurtzite GaN(0001)-(1x1) surfaces by ultraviolet  
photoemission. By extrapolating the leading peak, the VBM was determined at 2.4 eV above  
the strong structures, which are located in our theory around -1.0 eV. The difference in the positions of  
the VBM is 1.4 eV being much more than the accuracy of 0.05 eV which is assumed for  
this technique \cite{DiHo97}. The error in the experimental VBM
determination by
extrapolation appears to be critical. 
Beside the VBM the detailed comparison of calculated and measured
photocurrents allows us to determine the bulk band width to 7.2 eV.
Furthermore we identify emissions from a surface state, which is related to
the gallium adlayer. These emissions are a first hint for the
reliability of the used surface geometry and will be further analyzed with
the more surface sensitive off-normal photoemission in the next section.  	 
 
\subsection{Off-normal emission - theory} 
 
Priority of section (B) and (C) was given to analyze the electronic bulk
features from of normal emission spectroscopy.  
In this section we present theoretical results for off-normal emission along the  
$\overline{\Gamma M}$ and the $\overline{\Gamma K}$ direction.
In addition to the electronic structure we are now interested in the
geometric structure of the surface. In this context it seems necessary to
examine the real space origin of the photocurrent, which is done for two
examples before we consider the whole series.

\begin{figure}\centering 
\centerline{\psfig{figure=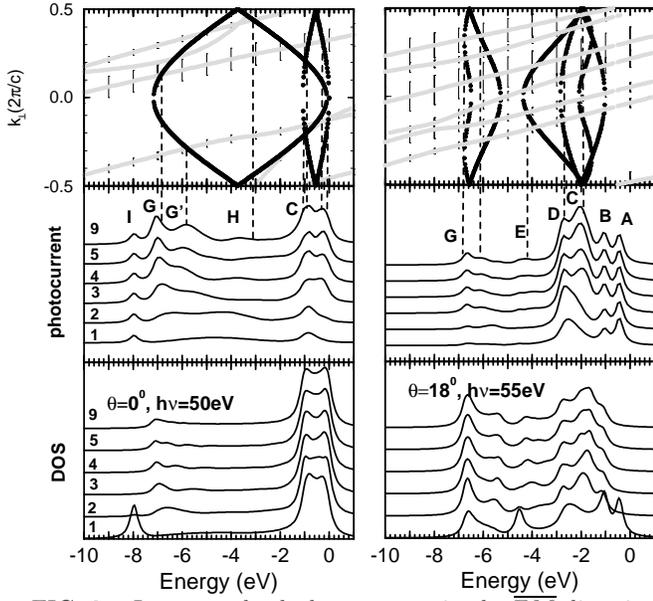,width=8.6cm}}
 \caption{\label{ganhexgam00ana} Layer resolved photocurrent in the
 $\overline{\Gamma M}$
 direction for $0^\circ$ and $18^\circ$ (middle part) together with the
 complex final bands (grey) and the initial bulk bands (dots) in the top
 panel. The final bands are shifted by the excitation energy and the 
 positions of direct transitions are indicated (vertical dashed lines).
 The bottom panel shows the
 layer-resolved density of states. The layer resolved photocurrent is 
 calculated for the number of layers as indicated, each layer consisting of
 four atomic planes (see text). }
\end{figure}

Figure \ref{ganhexgam00ana} shows two layer resolved spectra in the
$\overline{\Gamma M} $ direction. They are calculated for emission angles of 
$0^\circ$ (normal emission) and $18^\circ$ with photon
energies of 50 and 55 eV respectively. The numbers at the layer resolved
spectra indicate the number of layers which have been used in the sum of
Eq. 1, starting with the topmost layer.
The spectra are shown together with the density of states,
the bulk valence bands, and the complex  
final bands. The DOS is calculated for different ${\bf k_{\|}}$, referring to  
the plotted angles and binding energies, such that they can directly be compared with the photocurrents. 
Also the bulk bands are calculated taking into account the correct ${\bf k_{\|}}$. 
The complex final bands are shifted by the excitation energy onto the
valence band structure. For the photoemission spectra the light 
impinges in the $yz$ plane with an polar angle of $45^\circ$. 

For normal emission six peaks can be seen in the photocurrent 
(Fig. \ref{ganhexgam00ana}). The double peak (C) can be  
explained by direct transitions from the two topmost valence bands. The positions of the direct  
transitions are indicated by the dashed lines. As the peak at the lower
binding energy side becomes visible above the third layer (12 atomic
planes), the peak at the higher binding energy side shows also contributions
from the surface layer. 
These contributions are related to emissions from the  
nitrogen $p_z$ and $p_y$ orbitals which yield the largest matrix elements.  
Peak (H) is only a weak structure. It is explained by direct transitions into final bands which  
contribute less to the outgoing state. (G) and (G') are direct transitions from  
the lower valence bands into the two major final bands. Especially for (G) also  
emissions from the nitrogen $p_z$ orbitals from the third and fourth nitrogen layer have to be  
taken into account. Peak (I) is clearly related to the emissions from the gallium $s$ and  
nitrogen $p_z$ surface state located in the first atomic layers, as already explained in the  
section (B).  

\begin{figure}[h]\centering  
\centerline{\psfig{figure=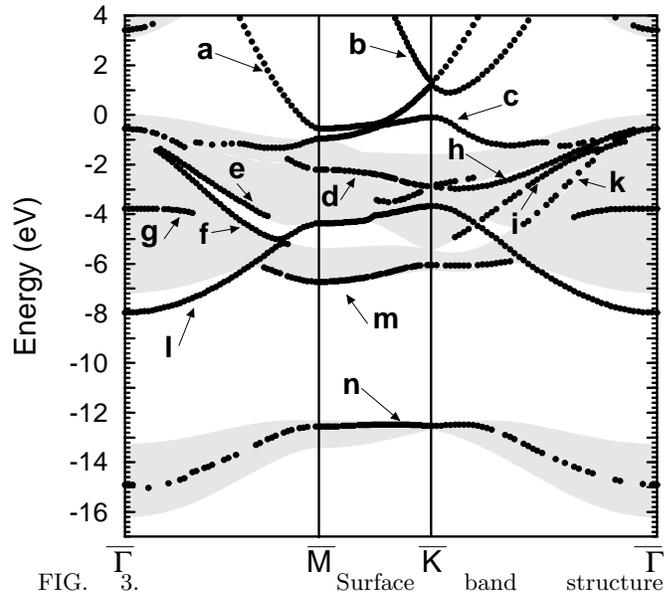,width=8.6cm}}        
 \caption{\label{ganoberband} Surface band structure of the 
 GaN(0001)-(1x1):Ga surface and the projected GaN bulk band structure 
 (shaded). For the calculation parameter set A according to Vogel
 {\it et al.}  
 \protect\cite{VoPo96} was used.}     
\end{figure}

Changing the emission angle to $18^\circ$ two emissions ((A) and (B)) appear, which
are due to the surface states (a) and (c) respectively (see Fig. \ref{ganoberband}). 
Both structures have their origin in the first atomic layers. 
The structures (C) and (D) are explained by direct transitions. In contrast
to the structure (C), which is visible above the third layer, structure (D)
shows an enhanced contribution from the surface layer (nitrogen $p_y$ and
$p_z$). These contributions are larger than those for the double peak (C) in
normal emission at -1.0 eV. 
The remaining peaks can be explained by direct transitions
with weak contributions from the nitrogen $p_z$ and $p_x$ orbitals of the
surface layers.  

Summarizing, we found an enhanced surface sensitivity for higher
angles. This regards the states within the gap, as well as resonant structures 
(e.g. (D) in Fig. \ref{ganhexgam00ana}). 
The enhanced surface sensitivity at off-normal angles is well known and has
been recently investigated
by one-step model calculations \cite{SoSc98}.

\begin{figure}\centering 
\centerline{\psfig{figure=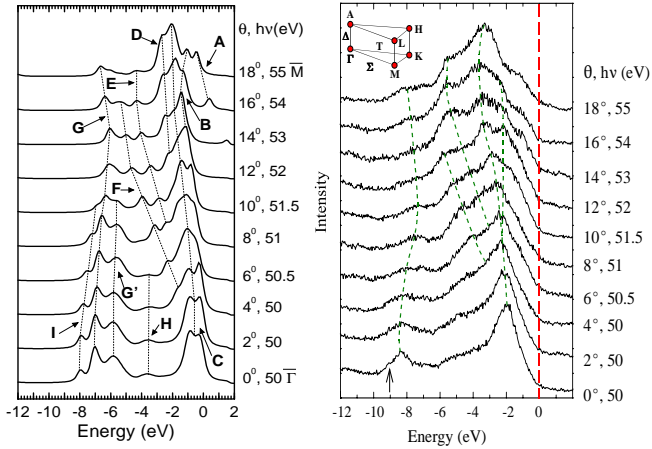,width=8.6cm}} 
 \caption{\label{ganhexgammdosphotoexp} Results for the $\overline{\Gamma M}$ direction.  
 On the left hand side the theoretical  
 photocurrents according to parameter set A, and on the right hand side experimental data  
 from Dhesi {\it et al.} \protect\cite{DhSm97}. The arrow indicates a weak
 shoulder to be compared with (I) (see text)} 
\end{figure}

In Fig. \ref{ganhexgammdosphotoexp} a series along the $\overline{\Gamma}\overline{M}$ 
direction is presented. The structure (A) can be related to the surface state (a) in the  
surface band structure (see Fig. \ref{ganoberband}) 
and is built up by emissions from the topmost nitrogen $p_z$ and  
gallium $p_z$ and $p_x$ orbitals. The emissions from structure (B) are related to the same  
orbitals and belong to the surface state (c). The structure (C) can be found
at all angles.  
Apart from direct transitions also emissions from the nitrogen $p_z$ and $p_y$ orbitals contribute.  
Especially, the structure (D) displays besides direct transitions the DOS of 
the first 8 atomic layers (see Fig. \ref{ganhexgam00ana} ). The  
emissions (E) and (F) can be explained by the surface resonances (e) and (f) (see Fig.  
\ref{ganoberband}) which frame a gap in the projected bulk band structure. The emissions are  
related to nitrogen $p_x$ orbitals of the first three layers with varying contributions from  
direct transitions. The dispersion of structure (G) follows the lower valence band
edge, and in addition to direct transitions, emissions from  
nitrogen $p_x$ and $p_z$ orbitals are responsible for this structure. 
Structure (I) belongs to the surface state (l), see Fig. \ref{ganoberband}.
The structure (G') is 
connected to direct transitions as can be seen in Fig. \ref{ganhexgam00ana}.

\begin{figure}\centering 
\centerline{\psfig{figure=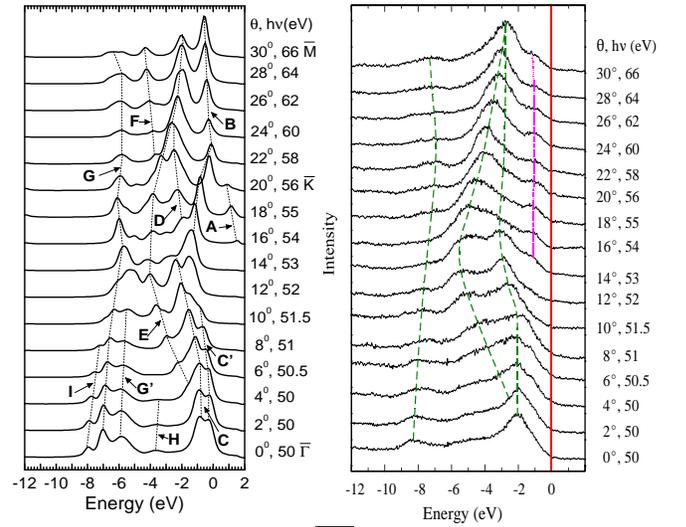,width=8.6cm}} 
 \caption{\label{ganhexgamkdosphotoexp} Results for the $\overline{\Gamma K}$ direction.  
 On the left hand side the theoretical  
 photocurrents according to parameter set A, and on the right hand side 
 experimental data 
 from Dhesi {\it et al.} \protect\cite{DhSm97}} 
\end{figure}

In Fig. \ref{ganhexgamkdosphotoexp} the theoretical photocurrents along the  
$\overline{\Gamma}\overline{K}$ direction are shown. The spectra are calculated for angles  
between $0^\circ$ and $30^\circ$, with photon energies between 50 and 66 eV. The light is $p$- 
polarized and incident along the $xz$- plane with an angle of $45^\circ$ with respect to the  
surface normal. Because of the high emission angles also the $\overline{K}\overline{M}$  
direction is reached.    
  
The theoretical spectra show a weak emission (A) for angles around $18^\circ$. This emission  
represents the surface band (a) between $\overline{\Gamma}$ and $\overline{K}$, which can  
be seen in Fig. \ref{ganoberband}. Structure (B) results from the surface band (c), which  
leaves at $14^\circ$ the projected bulk band structure.
The structures (C) and (C') can be explained by 
direct transitions from the topmost valence bulk band.
Additionally emission from the huge density of states from the N $p_x$
orbitals have to be taken into account (see the discussion of Fig.
\ref{ganhexgam00ana}).  
The emission (D) results from nitrogen $p_z$ orbitals below the first layer.
For angles below $14^\circ$ the structure (E) stems from
the surface resonance (k) (see Fig. \ref{ganoberband}). 
It consists of  
nitrogen $p_y$ orbitals and shows large dispersion to higher binding energies. 
Above $14^\circ$ (E) interferes with emissions from the surface resonance (l).
The structure  
(F) belongs to the surface resonance (l) consisting of nitrogen $p_z$ and gallium $s$ orbitals,  
with the main contributions from the nitrogen surface atoms. The remaining structures (H, G,  
G' and I) are explained as their counterparts for the $\overline{\Gamma}\overline{M}$  
direction.  

In both theoretical spectra we are able to identify emissions which are
related to the orbital composition of the topmost surface layers.
Moreover, also emissions from resonances show contributions from the
surface, as has been pointed out by the layer resolved photocurrent. If we are
able to identify these emissions in experimental data, clear fingerprints
for the assumed gallium adlayer structure would be indicated.

\subsection{Off-normal emission - comparison with experiment} 
 
In this section we compare our results in off-normal emission with experimental data of
Dhesi {\it et al.} \cite{DhSm97}. For comparing the spectra, it is also important
here to take into account the shift of 1.0 eV, which is necessary to adjust the VBM (see
section C).

In Fig. \ref{ganhexgammdosphotoexp} the spectra  for  
$\overline{\Gamma M}$ are shown. At low  
binding energies the experiment shows strong emissions, which are identified with the structure  
(C) in the theoretical spectra. Near -8.0 eV the experiment shows a    
structure, which disperses to lower binding energy for higher angles with decreasing intensity.
This behavior is also seen in the theoretical structure (G). The energetic
difference between the two experimental structures coincides in
$\overline{\Gamma}$ with the
theoretical valence band width of 7.2 eV and underlines the results from
section C.

For angles above $14^\circ$ the experimental data display a  
shoulder between -1.0 and -2.0 eV. This emission can be associated with  
structure (B) in theory. For $16^\circ$ and $18^\circ$ the shoulder becomes very broad, which  
may be attributed to the theoretical emissions (B) and (A). 
The theoretical orbital composition of  
these states is consistent with the results of Dhesi {\it et al.}, 
who examined the dependence of this shoulder on polarization and
contamination.
The theoretical spectra show
a further emission from a surface state (I). This emission   
might be identified with the high binding energy shoulder in the
experimental data near -9.0 eV.
Around -4.0 eV, the experiment  
shows two dispersing structures. These structures can be connected with the theoretical  
emissions (E) and (F) which appear to be weaker, however. 
Also the structure (G') is seen in experiment as a weak shoulder at low emission angles.  
Between -4.0 and -5.0 eV the experiment shows a structure ($\theta$ =  
$0^\circ$... $8^\circ$) not being marked which is related to the theoretical structure (H).  
Thus three surface states and several surface resonances can be identified in experiment.
Taking into account the influence of the topmost atomic layers to the
photocurrent (see section D) the coincidence of the theoretical and
experimental spectra confirm the used surface geometry. Moreover,
considering
the energetic shift of 1.0 eV, the energetic positions of the structures
confirm the used surface band structure.  

Further information can be reached with the results  
along the $\overline{\Gamma K}$ direction (Fig. \ref{ganhexgamkdosphotoexp}).   
For angles above $14^\circ$ experimental data show a structure near -1 eV.  
This structure can be associated with the emission (B) in the theoretical
curves  
which results from the nitrogen and gallium surface state. Compared to experiment theory  
heavily overestimates the intensity. Also, the theoretical band takes  off from the projected band  
structure background with rather strong dispersion already at $18^\circ$ (see also (c) in Fig.  
\ref{ganoberband}) delayed by $4^\circ$ with respect to experiment, where this band clearly  
appears already at $14^\circ$ with very low dispersion. This difference is a hint that the theoretical  
surface  
band (c) becomes free of the projected bulk band structure with smaller dispersion along  
$\overline{\Gamma K}$ and significant closer to the $\overline{\Gamma}$ point
than obtained in our surface band structure calculation. 
In the same energy range but for lower angles, a shoulder is marked in the spectra which is  
comparable with the structure (C') in theory.  
 
Near -2.0 eV experiment marks a  
structure, dispersing a little to higher binding energy for larger angles. This structure can be  
identified with the theoretical emission (C), visible for angles between $0^\circ$ and $12^\circ$.  
The experimental structure disperses from -2.0 eV to -3.0 eV,  
while the theoretical structure disperses from -0.9 eV to -2.4 eV at $12^\circ$. 
Above $14^\circ$ the experiment shows no peaks for this structure, but only shoulders. The  
theoretical curves show the structures (D) and (E), which explains for higher angles
these
shoulders and the adjacent  
experimental peaks on the higher binding energy side. 

Below $14^\circ$ structure (E) reproduces the dispersive experimental structure between
-2.6 and -5.6 eV. Different from experimental data, the emission (E)
is less pronounced and displays less dispersion (about 300 meV).
At still  
lower angles a weak emission (H) is seen, which is visible as a weak structure in the  
experimental data. Near the lower valence band edge, theory shows three structures (G,G' and  
I) which can be compared with the experimental structure around -8 eV. It displays  
similar behavior as the theoretical data with respect to dispersion and magnitude, though the  
experimental emissions are weaker. 
 
Summarizing, we are able to explain all observed experimental structures. The observed  
energetical positions underline our result in normal emission, namely that the band width  
is 7.2 eV and that the experimental valence band maximum has to be shifted for 1.0 eV to  
higher binding energies. In addition to direct transitions all emissions in off-normal emission  
are influenced by surface states and resonances, as has been verified by the
layer resolved photocurrent. This demonstrates the surface sensitivity of the  
experiment, misleading the mapping of valence bulk bands solely from off-normal  
measurements \cite{DhSm97}. Also, we are able to identify three surface emissions, which show the
same   
energetical and intensity behaviour in theory and experiment. They can be related to emissions 
from the topmost nitrogen $p_z$ and gallium $s$  
and $p_x$ orbitals. The theoretical dispersion is only slightly at variance with  
experiment. The emissions from the topmost atomic layers sensitively  
depend on surface structure and reconstruction. Thus the comparison between experimental  
and theoretical results confirms the reliability of the assumed  
theoretical surface model.    
 
\section{Conclusion} 
 
Photoemission spectra in normal and off-normal emission for the GaN(0001)-(1x1):Ga surface  
have been calculated within the one-step model. Normal emission spectra show emissions from  
a surface state near the lower valence band edge. It is identified by its 
energetic position different from the band edge and its varying intensity 
by inspection of the matrix elements.  
Furthermore, we demonstrate that a  
widespread  experimental method to determine the VBM by extrapolating the 
leading edge of the valence band spectra may fail by as much as 1.0 eV.  
Taking this into account all the experimental structures can be identified in close agreement with  
theory. Especially, the valence band width (7.2 eV) agrees with a 
LDA bulk band structure calculation of Vogel {\it et al.} whereas a GW
calculation of Rubio {\it et al.} differs by 0.8 eV.  
 
In off-normal emission, surface states near the upper valence band edge  can be identified 
and analyzed with  
respect to surface band structures. Several surface resonances are examined and verified by  
experimental data. Aggrement of surface specific properties in the theoretical and experimental   
photocurrents is seen as a proof of the used surface geometry. The surface is nitrogen  
terminated with a gallium adlayer. 
 
The experimental emissions are traced back by theory to their origin in band structure, electronic  
states, orbital composition and location in direct space. 
Thus the one-step model calculation is a powerful tool to yield essential insight into the bulk and  
surface  
electronic structure. In addition, it gives credit to the underlying surface geometry. This work  
stresses the necessity of  such a calculation for a reliable interpretation of experimental  
ultraviolet photoemission data in comparison with calculated band structures.

\section{Acknowledgments} 
Discussions with  Prof. M. Skibowski and Dr. L. Kipp are gratefully acknowledged.  
We thank Prof. K. E. Smith providing us the experimental figures. 
The work was supported by the BMBF, under contract no. 05 SB8 FKA7.

\onecolumn
\begin{table} 
 \Large 
\begin{tabular}[h]{cccccccccccc}  
&$K_{ss}$&$K_{sp}$&$K_{pp}$&$I_{s0}$&$I_{p0}$&$\tilde{I}_{s0}$&$\tilde{I}_{p0}$
& 
$I_{s1}$&$I_{p1}$&$\tilde{I}_{s1}$&$\tilde{I}_{p1}$ \\ \hline  
A&1.0&0.90&0.87&49.78&42.00&22.84&12.05&29.53&22.31&16.71&9.61  
\\ \hline  
B&1.0&0.844&0.844&54.66&39.94&20.49&7.76&24.28&21.19&13.82&4.71
\\   
\end{tabular} 
\caption{\label{eht}EHT-parameter for GaN. The first line (set A) is
adjusted to the band
structure of Vogel {\it et al.} \protect\cite{VoPo96}, while  
the second line (set B) is adjusted to Rubio {\it et al.}
\protect\cite{RuCo93}.  (0 means nitrogen
and 1 gallium,    
for the notation see Starrost {\it et al.}\protect\cite{StSc96}).} 
\end{table} 
\nopagebreak

\end{document}